\documentclass[12pt,preprint]{aastex}

\usepackage{emulateapj5}

\slugcomment{}

\shorttitle{Discovery of a Very Nearby Brown Dwarf to the Sun}
\shortauthors{Biller et al.}  

\begin{document}


\title{Discovery of a Very Nearby Brown Dwarf to the Sun: 
A Methane Rich Brown Dwarf Companion to the Low Mass Star SCR 1845-6357}



\author{B.A. Biller\altaffilmark{1}, M. Kasper\altaffilmark{2}, 
L.M. Close\altaffilmark{1}, W. Brandner\altaffilmark{3}, $\&$ 
S. Kellner\altaffilmark{4}}

\email{bbiller@as.arizona.edu}

\altaffiltext{1}{Steward Observatory, University of Arizona, Tucson, AZ 85721, USA,bbiller@as.arizona.edu, lclose@as.arizona.edu}
\altaffiltext{2}{European Southern Observatory, Karl-Schwarzschild-Strasse 2, 
85748 Garching, Germany, mkasper@eso.org}
\altaffiltext{3}{Max-Planck-Institut f\"ur Astronomie, K\"onigstuhl 17, 69117 
Heidelberg, Germany, brandner@mpia.de}
\altaffiltext{4}{W.M. Keck Observatory, 65-1120 Mamalahoa Hwy., Kamuela, HI 
96743, USA, skellner@keck.hawaii.edu}




\begin{abstract} 

We present VLT/NACO SDI images of the very nearby star
SCR 1845-6357 (hereafter SCR 1845).  SCR 1845 is a recently discovered 
\citep{ham04}
M8.5 star just 3.85 pc from the Sun \citep{hen06}.  Using the capabilities
of the unique SDI device, we discovered a substellar companion 
to SCR 1845
at a separation of 4.5 AU (1.170''$\pm$0.003'' 
on the sky) and fainter by 3.57$\pm$0.057 mag in the 1.575 $\mu$m SDI filter.
This substellar companion
has an H magnitude of 13.16$^{+0.31}_{-0.26}$ (absolute H magnitude of 
15.30$^{+0.31}_{-0.26}$), making it likely the brightest mid-T dwarf known.  
The unique Simultaneous Differential Imager 
(SDI) consists of 3 narrowband filters placed around
the 1.6 $\mu$m methane absorption feature characteristic of T-dwarfs
(T$_{eff}$ $<$ 1200 K). 
The flux of the substellar companion drops by a factor of 2.7$\pm$0.1
between the SDI F1(1.575 $\mu$m) filter and the SDI F3(1.625 $\mu$m) filter, 
consistent with strong methane absorption in a substellar companion. 
We estimate a spectral type of T5.5$\pm$1 for the companion
based on the strength of this methane break.  
The chances that this object is a background
T dwarf are vanishing small -- and there is no isolated 
background T-dwarf in this part of the sky according to 2MASS.
Thus, it is a bound companion, hereafter SCR 1845-6357B.
For an age range of 100 Myr - 10 Gyr and spectral type range
of T4.5-T6.5, we find a mass range of 9 - 65 M$_{Jup}$
for SCR 1845B from the \citet{bar03} COND models.  
SCR 1845AB is the 24th closest stellar system to the Sun (at 3.85
pc); the only brown dwarf system closer to the Sun is the binary
brown dwarf Eps Indi Ba-Bb (at 3.626 pc).  In addition, this is the first
T-dwarf companion discovered around a low mass star.

\end{abstract}


\section{Introduction}
 
After decades of 
little change in the number of known stellar systems within 5 pc,  
numerous previously unknown low mass stars have recently been 
discovered in the solar neightborhood \citep[e.g.~][]{sch03,hen04,ham04}.
Because they are extremely nearby and intrinsically low luminosity, 
these objects are ideal targets to search for low mass companions, 
since even a close companion will appear reasonably separated on the sky.
For example
during commissioning of the Simultaneous Differential Imager (SDI) at the
VLT \citep{clo05a,len04}, one such object,
$\epsilon$ Indi B was resolved into a binary T dwarf \citep{mcc03}.
$\epsilon$ Indi Ba,Bb are the closest brown dwarfs to the Earth 
(3.626$\pm$0.009 pc).
At a distance of 3.85$\pm$0.02 pc \citep{hen06}, the recently discovered 
M8.5 star SCR 1845-6357 is just slightly further away 
than $\epsilon$ Indi Ba,Bb \citep{ham04} and is the 24th closest
stellar system from the Earth.  We report the discovery of SCR
1845-6357B (hereafter SCR1845B), a methane rich 
substellar companion to this star.  This companion object 
is the third closest brown dwarf 
to the Sun.  It is also 
the only example of a T dwarf companion to a low mass star and 
one of the tightest known brown dwarf companions to a star.

\section{Observations and Data Reduction}

Data were taken on the night of 2005 May 28 (UT)\footnote[1]{Based on
observations collected at the European Southern Observatory, Paranal,
Chile through proposal 075.C-0357(A)} at the 8.2m VLT-UT4 
with the unique Simultaneous Differential Imager (SDI) in 
the facility AO system NACO \citep{len03,rou03}.
To guide on this faint red object, 
the infrared wavefront sensor (IRWFS) was used with the ``K''
dichroic, sending all of the K band light to the WFS.  SDI can be 
used to calibrate and remove the ``speckle noise'' in AO images, while 
also isolating the light from a substellar methane companion 
from the starlight.  This method was 
pioneered by \citet{rac99}, \citet{mar00}, \citet{mar02}, and 
\citet{mar05}.
It exploits the fact that all cool (T$_{eff}$ $<$1200 K) 
substellar objects have strong CH$_4$ (methane) absorption 
redwards of 1.62 $\mu$m in the H band infrared atmospheric window 
\citep{bur01,bur03}.  The NACO SDI device obtains four 
images of a star simultaneously through three slightly different
narrowband filters
\citep[sampling~both~inside~and~outside~of~the~CH$_4$~features~
--][]{clo05a,len04}.  These images are then differenced.  This
subtracts out the halo and speckles from the bright star to reveal any
substellar methane objects orbiting that star.  Since a substellar
methane object will be brightest in one filter and absorbed in the
rest, while the star is bright in all three, a difference can be
chosen which subtracts out the star's light and reveals the
companion's light.  Thus, SDI also helps eliminate the large contrast
difference between the star and substellar companions
\citep{clo05a,len04,len05}.  The SDI device has already produced a
number of important scientific results: the discovery of AB Dor C
\citep{clo05b} -- the faintest companion ever discovered within 0.16''
of a star, detailed surface maps of Titan \citep{har04}, the discovery
of the binarity of $\epsilon$ Indi Ba-Bb, the nearest binary brown
dwarf \citep{sch03,mcc03}, and evidence of orbital motion for Gl 86B,
the first known white dwarf companion to an exoplanet host star
\citep{mug05}.  Using the SDI device provides a marked advantage over
single band imaging even in situations where the contrast difference
between star and companion is not large -- images in the 3 different
SDI filters immediately provide spectral information about any
substellar candidate, particularly regarding the amount of CH$_4$
present.

SCR 1845 was observed for 15 minutes (with 3$\times$30 s subimages taken 
at 5 different dither positions) at a position angle of 0$^{\circ}$
and 15 minutes at a position angle of 22$^{\circ}$.  A base integration time 
(DIT) of 30 s was used and subimages were medianed.
Observing the object at different roll angles allows 
us to immediately confirm if an object is real -- an instrumental feature
should not rotate with a change of 
rotator angle; however, a real object on the sky 
should appear to rotate by the change in 
rotator angle.  Data were sky-subtracted, flat-fielded, and bad-pixel masked. 
Each data frame was then aligned to a master frame
using the IRAF task xreg.  After alignment, all frames were median combined.
As a comparison, the data were also reduced using a custom IDL SDI 
pipeline which performs basic data reduction tasks and also precisely aligns
images taken in each of the filters using a custom shift and subtract 
routine \citep{bil05}.

\section{Results and Discussion}

\begin{figure}
 \includegraphics[angle=0,width=\columnwidth]{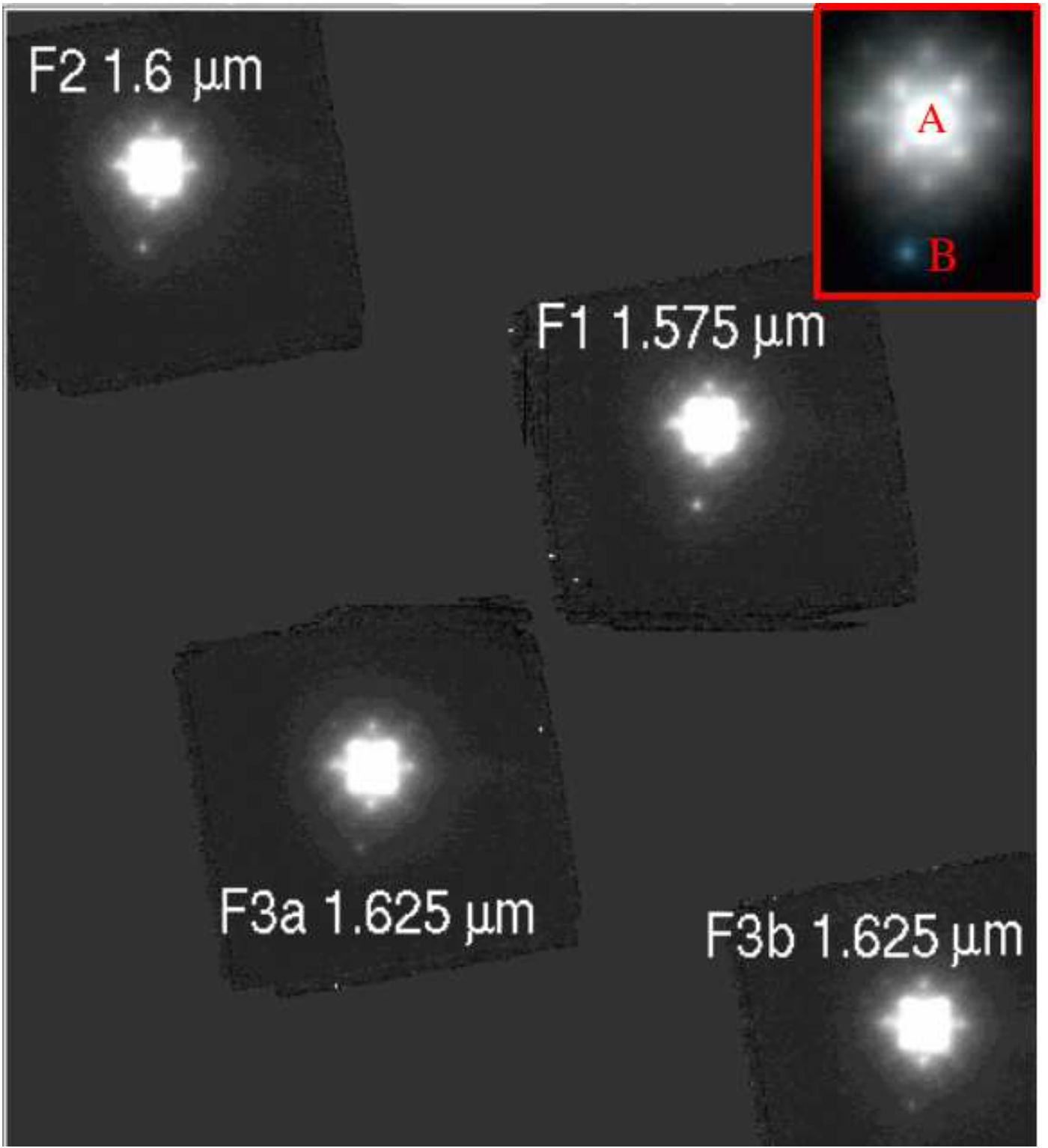}
\caption{An SDI image of SCR 1845.  This 15 minute long
image was taken at a position angle of 0$^{\circ}$ and was reduced
using a custom IDL pipeline and the IRAF xreg tool.  A substellar
companion appears at a separation of 1.170''$\pm$0.003'' (4.5 AU at
3.85 pc) from the primary and a position angle of
170.20$\pm$0.13$^{\circ}$ in each of the 4 SDI filters.  The
platescale is (0.01725''$\pm$0.00025'')/pix \citep{nie05}.  The
companion appears brightest in the F1 filter (out of the CH$_4$
absorption) and drops by a factor of 2.7 in the F3 filter (inside the
CH$_4$ absorption), consistent with a T5.5 dwarf spectral type.  North
is up and east is to the left.  {\bf Upper Right Inset:} Three color
image of SCR 1845 A and B generated from the SDI filter images
(blue=1.575 $\mu$m, green=1.600 $\mu$m, red=1.625 $\mu$m).  The
substellar companion appears blue in this image.  This image was
created with a log10 stretch and each filter is equally weighted.
Note how similar in color (white) each of the PSF speckles are for the
M8.5, while the faint companion SCR 1845B is considerably bluer due to
strong CH$_4$ absorption.  The structure in the PSF is typical of the
NACO IR WFS for a faint guide star such as SCR 1845A. }
\label{fig:reduction0}
\end{figure}

\begin{figure}
  \includegraphics[angle=0,width=\columnwidth]{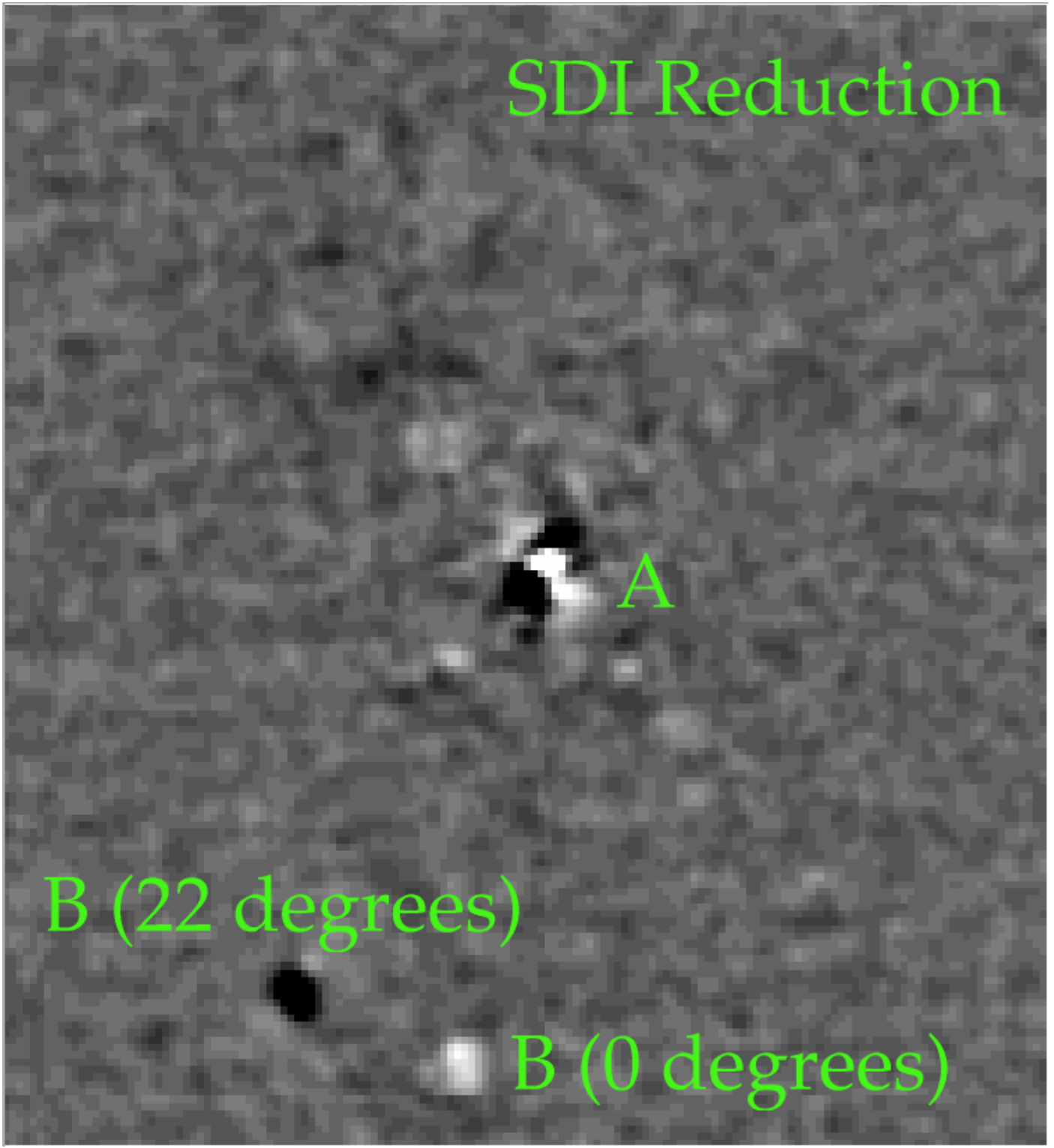} \\      
\caption{
Images of SCR 1845 using the SDI device and reduced using a custom SDI
pipeline \citep{bil05}.  This 30 minute long image was taken at
position angles of 0$^{\circ}$ (white) and 22$^{\circ}$ (black).
Datasets from each roll angle were subtracted from each other and
smoothed with a 1 pixel FWHM gaussian.  A substellar companion appears
at a separation of 1.17'' from the primary in each of the 4 SDI
filters. Note that the speckles from the M8.5 are almost totally
removed.  With the high contrasts achievable by SDI, a methane object
like SCR 1845B ($\Delta$H=4.2 mag) could have been detected at
10$\sigma$ 10$\times$ closer in at a separation of only $\sim$0.1''. }
\label{fig:3colorSDI}
\end{figure}

Reduced data for the 0$^{\circ}$ dataset are presented 
in Fig.~\ref{fig:reduction0}.  SCR 1845B appears at 
a separation of 1.170''$\pm$0.003'' and at a position angle of 
170.20$^{\circ}\pm$0.13$^{\circ}$
from the M8.5 primary in all four of the SDI filters
and rotates by 22$^{\circ}$ (as expected) between datasets.
A three color image 
generated from the SDI filter images is presented in 
Fig.~\ref{fig:3colorSDI}. 
For comparison, an image reduced using the SDI pipeline \citep{bil05}
is also presented in Fig.~\ref{fig:3colorSDI}.  While SCR 1845B is 
far from the primary and easily detected, we would also be capable of 
detecting similar or lower mass companions closer to the primary 
for this system (down to 0.1'' separations).

\citet{dea05} measured a trigonometric parallax 
to SCR 
1845-6357 of 282$\pm$23 mas, corresponding to a distance of 3.5$\pm$0.3 pc.
With additional epochs of observation, 
\citet{hen06} provide an updated distance of 3.85$\pm$0.02 pc.
Thus, the candidate substellar object lies 4.50$\pm$0.02 AU from its primary.
The candidate object is 3.57$\pm$0.057 mag fainter than the primary 
in the F1(1.575 $\mu$m) filter (all photometry performed
with the IRAF DAOPHOT PSF fitting package).  

\subsection{Spectral Type}

\begin{figure}
 \includegraphics[angle=0,width=\columnwidth]{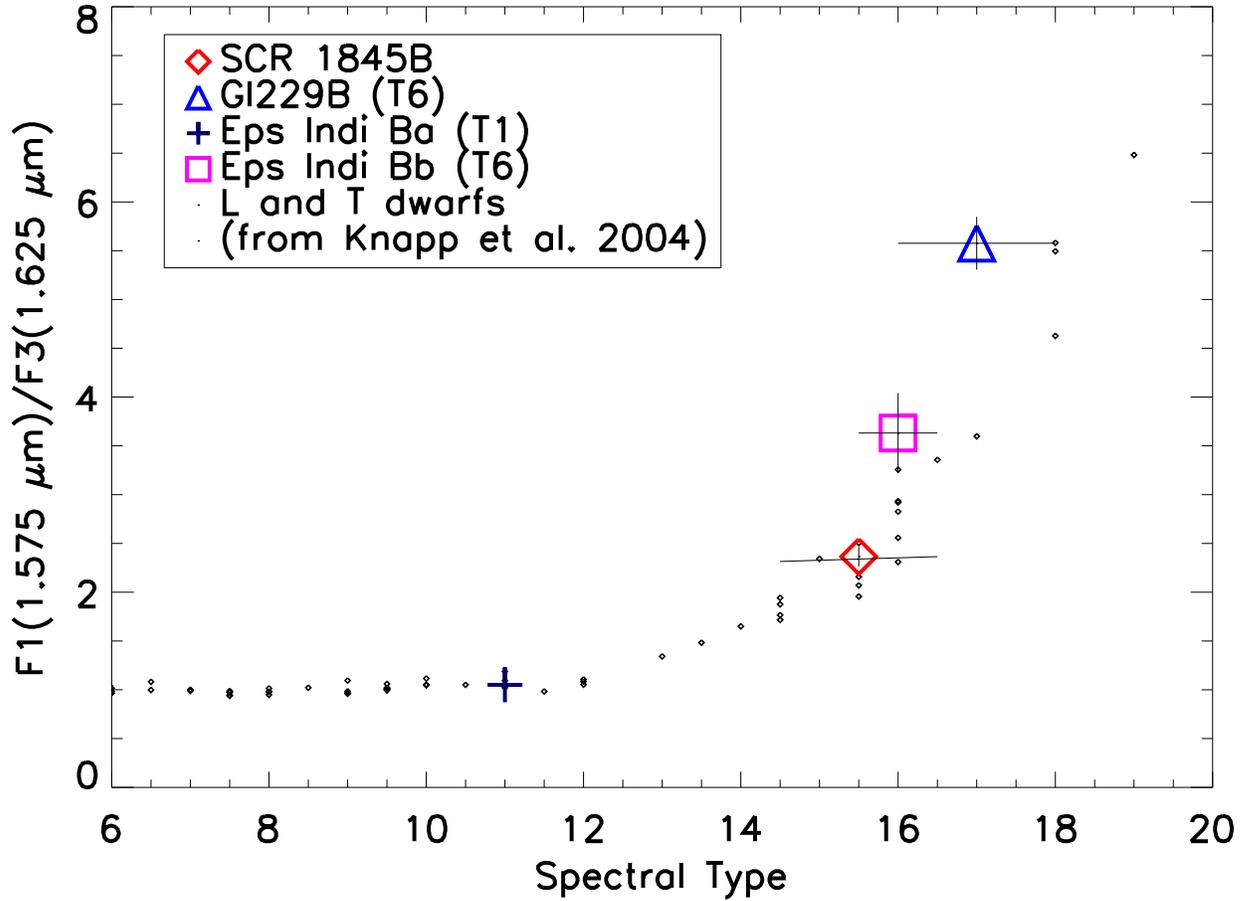}
\caption{SDI methane spectral indices for SCR 1845B and the T dwarfs
Gl 229B, $\epsilon$ Indi Ba, and $\epsilon$ Indi Bb.
We plot numerical spectral types on the x-axis; a numerical
type of 8 corresponds to a L8 spectral type, a numerical type of 16 
corresponds to a T6 spectral type, etc.
As a comparison, SDI methane spectral indices calculated from spectra for 94
L and T dwarfs \citep[spectra~from~][]{kna04} are overplotted
as small dots. 
SCR 1845B, Gl 229B, and $\epsilon$ Indi Bb (T6) show strong methane indices, 
whereas $\epsilon$ Indi Ba (T1) is relatively constant in flux 
across the SDI filters and has a much lower methane index.  \citet{geb02}
note that Gl 229B has an anomalously high
methane index for its spectral type.  While \citet{geb02} find an 
overall spectral type of T6$\pm$1 for Gl 229B, they assign Gl 229B a 
spectral type of T7 based on this methane index.  Note that the 
spectral indices for SCR 1845B are only consistent with spectral
types of T4.5-T6.5.}
\label{fig:fluxes}
\end{figure}

The candidate object appears brightest in the F1(1.575 $\mu$m) filter,
slightly fainter in the F2(1.6 $\mu$m), and then drops by a factor of
2.7$\pm$0.1 between the F1(1.575 $\mu$m) and F3(1.625 $\mu$m) filters.
The spectral signature of this dropoff is consistent with methane
absorption in the atmosphere of a substellar object \citep{geb02}.
Previous observations of the T6 spectral type brown dwarf $\epsilon$
Indi Bb \citep{mcc03} with the SDI device found that the flux of
$\epsilon$ Indi Bb also dropped by a similar factor between the
F1(1.575 $\mu$m) and F3(1.625 $\mu$m) filters (see
Fig.~\ref{fig:fluxes}).  To determine an accurate spectral type for
SCR 1845B, we define an SDI methane spectral index calculated from our
SDI F1(1.575 $\mu$m) and F3(1.625 $\mu$m) filter images
\citep[similar~in~concept~to~the~methane~spectral
index~defined~by~][]{geb02}.  The SDI device measures the location and
strength of the 1.6 $\mu$m methane absorption break, which is a
principle spectral feature used to determine spectral types for T
dwarfs -- this SDI methane index should be sufficient to estimate an
accurate spectral type for this object.  The SDI methane spectral
index is defined as: index($\frac{F1}{F3}$) =
$\frac{\int^{\lambda_2}_{\lambda_1} S_{\lambda} F1({\lambda})
d\lambda}{\int^{\lambda_4}_{\lambda_3} S_{\lambda} F3({\lambda})
d\lambda}$

Each SDI filter was manufactured by 
Barr Associates to have a precise bandwidth of  
0.025 $\mu$m, so the wavelength
intervals in the numerator and denominator have the same length for the 
SDI methane index.

We only possess SDI data on a limited number of T dwarfs (this object,
Gl 229B, $\epsilon$ Indi Ba (T1), $\epsilon$ Indi Bb (T6)).  
In order to compare SCR 1845B to 
a wider range of L and T dwarf objects we calculated these same SDI 
spectral 
indices from spectra of 56 L dwarfs and 35 T dwarfs \citep{kna04}.
Spectra for these objects were obtained from 
Sandy Leggett's L and T dwarf archive\footnote[2]{http://www.jach.hawaii.edu/$\sim$skl/LTdata.html}.  In order to make an accurate comparision, 
SDI filter transmission curves were convolved into these
calculations.  Since we have full spectral data for these objects, 
we also calculated the 1.6 $\mu$m methane spectral
index defined by \citet{geb02}, which were found to be similar to our 
SDI methane spectral indices.  In Fig.~\ref{fig:fluxes}, 
SDI methane spectral indices are plotted for SCR 1845B, the T dwarfs
Gl 229B, $\epsilon$ Indi Ba, $\epsilon$ Indi Bb, 
and 94 other L and T dwarfs.
SCR 1845B appears to have a noticeable methane break with somewhat 
lower indices than the T6 dwarfs Gl 229B and $\epsilon$ 
Indi Bb.  However, \citet{geb02}
note that Gl 229B has an anomalously high
methane index for its spectral type and assign a large 
uncertainty to Gl 229B's spectral type -- T6$\pm$1.
For our SDI methane indices, SCR 1845B has
spectral indices similar to that of T4.5-T6.5 dwarfs.
Thus, we determine an initial spectral type of T5.5$\pm$1 for SCR 1845B.   

\subsection{H magnitude}

To determine an accurate H magnitude, the spectra of both the primary
and secondary components of SCR 1845 must be taken into account.  The
M8.5 primary is extremely red -- and will appear brighter in the H
band than in our blue F1 band.  Additionally, the T5.5$\pm$1 companion
is blue compared to the primary and will appear brighter in the F1
band than in the H band.  To convert from our F1 filter magnitudes
into calibrated H band magnitudes we must calculate the H band
magnitude offsets for the M8.5 primary star and the T4.5-T6.5
companion (Offset$_M$ and Offset$_T$ respectively): $ \Delta H = H_T -
H_M = (Offset_T + F1_T) - (Offset_M + F1_M) = (Offset_T - Offset_M) +
\Delta F1 $

Using the spectrum of the star VB10 (an M8 template, and thus a reasonable
approximation of an M8.5 spectrum), an H transmission
curve, and our F1 filter transmission curve, we calculate a magnitude 
offset of Offset$_M$=-0.12$\pm$0.08 mag.
Assuming spectral types of T4.5-T6.5, we can perform a similar
calculation for the companion.  Offsets were calculated for 
15 objects with spectral types of T4.5-T7 
\citep[spectra~from~][]{kna04}, then averaged together by spectral
type to derive an average offset for each spectral type. 
For instance, for a T5 companion, Offset$_{T5}$ = 
0.5$\pm$0.05.  For a T6 companion, Offset$_{T6}$ = 0.6$\pm$0.07.
 Magnitudes in the H filter for both 
primary and candidate object are presented in Table~\ref{tab:properties}.
Uncertainties are provided for a companion
spectral type of T5.5$\pm$1.  Background T4-T7 dwarfs possess absolute 
H magnitudes of $\sim$14.5-16.0 \citep{burg04}, so our calculated 
absolute H magnitude of 15.23$^{+0.31}_{-0.26}$ for SCR 1845B is 
quite reasonable.

\subsection{Likelihood of Being a Bound Companion and T Dwarf Number Densities}

This object has not been observed at multiple epochs
with the SDI device so we must consult other sources to 
determine if it is indeed truly bound, i.e. shares a common-proper 
motion with its primary.  SCR 1845 possesses a 
large proper motion of $\sim$2.5''/ year.  A bound companion 
would possess a similar proper motion whereas a background object would
appear to stay in the same spot on the sky.  On 2000 January 1, 
SCR 1845A had an RA of 18$^h$45$^m$05.2'' and DEC of 
-63$^{\circ}$57'47.355'' \citep[J2000,][]{dea05}.  Taking
proper motion into account, during the 2005 May 28 SDI observations, 
SCR 1845A therefore had an RA of 18$^h$45$^m$07.21'' and DEC
of -63$^{\circ}$57'43.586''.  
SCR 1845B (1.17'' separation at a PA of 170.2$^{\circ}$)
had an RA of 18$^h$45$^m$07.33'' and DEC of -63$^{\circ}$57'42.786''
during the 2005 May 28 SDI observations.  If the faint companion
is actually a background T dwarf, this RA and DEC should be reasonably correct
for other epochs of observation.
Checking the 2MASS
point source catalog (2MASS images 
taken 2000 May 29), we found no objects within 20'' of this position and no 
objects with T dwarf colors in this part of the sky.  Hence, it is 
impossible that this object is a background T-dwarf and it is highly 
likely to be a bound companion.

In the last few years, 3 new T dwarf companions have been 
discovered within 6 pc of the Sun 
--SCR 1845B and $\epsilon$ Indi Ba-Bb.  All three of the nearest T dwarfs 
are bound companions to stars.   
Combining these three new T dwarf companions with Gl 229B (5.8 pc) and 
Gl 570D (5.9 pc), we find
a number density of T dwarf companions of 5.5$\times$10$^{-3}$
pc$^{-3}$ within 6 pc of the Sun.
In contrast, only two isolated, field T dwarfs 
\citep[2MASS~J0415~and~2MASS~J0937~from][]{vrb04}
are known within 6 pc of the sun, leading to a rough number density of 
field T dwarfs of $\sim$2$\times$10$^{-3}$pc$^{-3}$ in this volume.  
Granted, this number of field dwarfs is somewhat incomplete
since there may be nearby T dwarfs without accurate trigonometric 
parallaxes, but the number is unlikely to change by more than a factor
of 2 since this population of bright T-dwarfs is very well studied.  
The existence of 5 brown dwarfs
within binary systems $<$ 6 pc of the Sun suggests that the number 
density of T dwarfs in binary systems may be higher than that of isolated, 
single T dwarfs.  This may be difficult to expain with ``ejection'' theories
of brown dwarf formation \citep{rei02}.

\subsection{Mass Estimate for SCR 1845B}

While the distance to SCR 1845 is well known, the age of the system is 
unconstrained.  Ages between 100 Myr and 10 Gyr are all 
plausible for this system at this time.  
Future observations of lithium absorption
might constrain the age of this system or at least rule out very young ages.
However, with a v$_{tan}$=41 km/s it is 
unlikely that this system is very young. 
Using the \citet{bar03} COND models with this age range,
an absolute H mag range of 15.1 to 15.6, and spectral types of T4.5-T6.5 
(T$_{eff}$ $\sim$ 850 K $\pm$ 100 K),  
we find a mass range of 9 - 65 M$_{Jup}$.  While SCR 1845B is 
clearly substellar at any age, 
the uncertainty in the age of this system means that
we cannot derive an unambiguous mass for this object from the COND models.
However, since this object is so close to its primary 
(currently $\sim$4.5
 AU), orbital motion should be 
evident within a few years.  Both the primary and secondary mass can
be measured accurately within a decade, making SCR 1845B a key T-dwarf 
mass-luminosity calibrator.

\section{Conclusions}

SCR 1845B is the brightest mid-T dwarf yet discovered.
In addition, it is the first T dwarf 
companion found around a low mass star.
At only $\sim$4.5 AU from its primary, it is one of the 
tightest known brown dwarf companions to a star 
and is a further piece of evidence that the brown dwarf desert does not 
exist for companions to very low mass stars \citep[][]{clo03,giz03}.
Both the primary and secondary mass can be accurately measured
within a decade.

\acknowledgements

BAB is supported by the NASA GSRP grant NNG04GN95H and NASA Origins
grant NNG05GL71G.  LMC is supported in part by an NSF CAREER award.
This research has made use of data products from the SuperCOSMOS Sky
Surveys at the Wide-Field Astronomy Unit of the Institute for
Astronomy, University of Edinburgh and from the Two Micron All Sky
Survey, a joint project of the University of Massachusetts and
IPAC/Caltech, funded by NASA and the NSF.  We thank Sandy Leggett for
providing electronic L and T dwarf spectra via her website.  We thank
Ralf-Dieter Scholz for alerting us to the updated parallax for SCR
1845A and Todd Henry for providing this parallax in advance of
publication.  We thank John Gizis for refereeing this paper and for
useful suggestions.

\clearpage

\begin{table}\def~{\hphantom{0}}
  \begin{center}
  \caption{SCR 1845 Photometry}
  \label{tab:properties}
  \begin{tabular}{ccccc}\hline
Spectral Type & H mag (primary)$^*$ & $\Delta$F1 & H (companion) & absolute H \\ \hline
T5.5$\pm$1 & 8.967 $\pm$ 0.027& 3.57 $\pm$ 0.057 & 13.16$^{+0.31}_{-0.26}$ & 
15.30$^{+0.31}_{-0.26}$ \\ \hline
~$^*$ from 2MASS & point source catalog & \\
  \end{tabular}
  \end{center}
\end{table}

\clearpage

\clearpage

\clearpage

\end{document}